\documentclass{JINST}
\usepackage{graphicx}
\usepackage{dcolumn}
\usepackage{etoolbox, ifpdf}
\usepackage{relsize}
\usepackage{soul}
\usepackage{bm}
\usepackage{booktabs}
\usepackage{subcaption}

\title{Characterization of the International Linear Collider
damping ring optics}
\author{J. Shanks, D.L. Rubin, and D. Sagan\\
CLASSE, Cornell University, Ithaca, New York 14853, USA\\
\email{js583@cornell.edu}}

\abstract{
    A method is presented for characterizing the emittance dilution
    and dynamic aperture for an arbitrary closed lattice that includes
    guide field magnet errors, multipole errors and misalignments.
    This method, developed and tested at the Cornell Electron Storage
    Ring Test Accelerator (CesrTA), has been applied to
    the damping ring lattice for the International Linear Collider (ILC).
    The effectiveness of beam based emittance tuning is limited by beam
    position monitor (BPM) measurement errors, number of corrector
    magnets and their placement, and correction algorithm. The specifications
    for damping ring magnet alignment, multipole errors, number of BPMs,
    and precision in BPM measurements
    are shown to be consistent with the required emittances and
    dynamic aperture. The methodology is then used to determine the
    minimum number of position monitors that is required to achieve
    the emittance targets, and how that minimum depends on the location
    of the BPMs. Similarly, the maximum tolerable multipole errors are
    evaluated. Finally, the robustness of each BPM configuration with respect
    to random failures is explored.
}


\begin{document}



\section{Introduction}
    The International Linear Collider (ILC) design utilizes damping
    rings to cool beams delivered by the electron and positron
    sources before transferring to the main linacs \cite{ilc:tdr}. Three of the
    primary requirements of the baseline damping rings are: 1) they must
    accept an injected bunch from the positron source with
    normalized phase-space amplitude $0.07~\textrm{m}\cdot
    \textrm{rad}$ and $\delta_E/E = 0.75\%$; 2) the beams must
    be cooled to an equilibrium zero-current geometric vertical
    emittance $\le 2~\textrm{pm}$; and 3) the damping time must
    be short enough to provide fully damped bunch trains at a repetition
    rate of 5Hz.

    The above requirements must be met in a real machine that includes
    magnet misalignments, guide field multipole errors
    (both systematic and random), and beam position monitor (BPM) measurement
    errors. Emittance tuning will be essential to
    achieve the target zero-current emittance. Quadrupoles and corrector
    magnets will necessarily be independently powered, allowing for localized
    corrections through the use of beam-based measurements.

    The damping rings cannot be characterized with respect to the exact
    set of errors they will have, as the rings are not yet built.
    For the purposes of this study, alignment
    errors in the damping rings are assumed to be randomly distributed
    with amplitudes dictated by survey tolerances. By simulating a large
    number of lattices with random distributions of errors at the appropriate
    levels, the damping rings can be characterized with a statistical analysis
    of the likelihood that the required emittances and dynamic aperture will be
    achieved. A configuration is deemed acceptable if 95\% of the
    randomly-misaligned and corrected lattices meet the required vertical
    emittance and dynamic aperture, as per the Technical Design Report
    specifications.

    Several methods for optics correction have been
    developed at storage ring light sources.
    By far the most widespread method is
    response matrix analysis (RMA), and specifically, orbit
    response matrix (ORM) analysis \cite{SLACPUB:9464}.
    The method involves taking difference orbits
    where corrector strengths are varied. Both the
    Swiss Light Source and Australian Synchrotron have
    demonstrated the capability of correcting vertical emittance
    below 1~pm using ORM \cite{NIMA:SLS,PRSTAB14:012804}, below the
    required vertical emittance for the ILC damping rings. However,
    the time required for ORM data acquisition
    scales linearly with the number of correctors in the ring.
    For large rings such as the proposed ILC damping rings with over
    800 steering correctors, acquiring difference orbits becomes a
    time-consuming process.

    Another increasingly-common class of optics correction algorithms
    is based on turn-by-turn BPM measurements, where the beam is
    pinged with a short-duration kick and allowed to oscillate
    freely \cite{castro_betatron_1993, PhysRevSTAB.16.012802}.
    Typically this data is then processed as a resonance-driving
    term (RDT) correction \cite{franchi_vertical_2011}.
    This correction technique has the benefit that
    correction times are roughly independent of the size of the ring.
    For the ILC this represents a considerable savings in time. While
    this method has shown promise, there are limitations.
    The beam decoheres due to amplitude dependent tune shift,
    limiting the number of useful turns of data.
    The decoherence may be partially mitigated
    by reducing the chromaticity to near zero, however measurements
    are still limited to a few thousand turns. As mentioned in
    \cite{PhysRevSTAB.16.012802}, the resolution of the betatron
    phase measurement scales roughly inverse with the number of
    turns ($1/N$). The limitations on the number of turns in a data
    set therefore directly correspond to a limitation on accuracy of
    betatron phase measurements.

    A turn-by-turn-based alternative to RDT has been developed for use
    at the Cornell Electron Storage Ring (CESR) for CESR Test Accelerator
    program (CesrTA) \cite{ICFABDNL50:11to33,PhysRevSTAB.17.044003}.
    The CESR betatron phase and coupling measurement
    relies on phase-locking turn-by-turn kickers to resonantly
    excite a single bunch to an amplitude of
    several millimeters, thus avoiding limitations due to
    bunch decoherence and allowing for data sets of over 40,000 turns.
    The effectiveness of the betatron phase/coupling correction procedure
    has been evaluated at CesrTA through measurements and simulations,
    demonstrating good agreement and validating our understanding of the
    limitations of optics correction.

    In this paper, the method for modeling optics correction developed at
    CesrTA is used to characterize the ILC damping rings. Several scenarios
    are evaluated, including the possibility of reducing the total BPM count
    and relaxing the constraints on guide field multipole errors, with respect
    to the baseline design. The effect of random BPM failures is analyzed to
    benchmark the robustness of each of the proposed BPM distributions.

    The characterization is comprised of two parts:
    generating a randomly misaligned lattice and simulating the
    optics correction procedure, and analyzing the dynamic aperture of the
    corrected lattice as a function of guide field multipole
    errors. The software used in all analyses is based on the
    \texttt{Bmad} accelerator code library \cite{NIMA558:356to359}.

    Only single-particle dynamics from the beam optics are
    considered in the work presented here. The
    reported emittances therefore represent the zero-current limit.
    Other current-dependent sources of emittance dilution, such as intra-beam
    scattering (IBS), electron cloud, and fast-ion instability
    will increase the vertical emittance beyond this lower limit.

\section{DTC04 Lattice}
\begin{figure}[tbh]
    \centering
        \includegraphics[width=4 in]{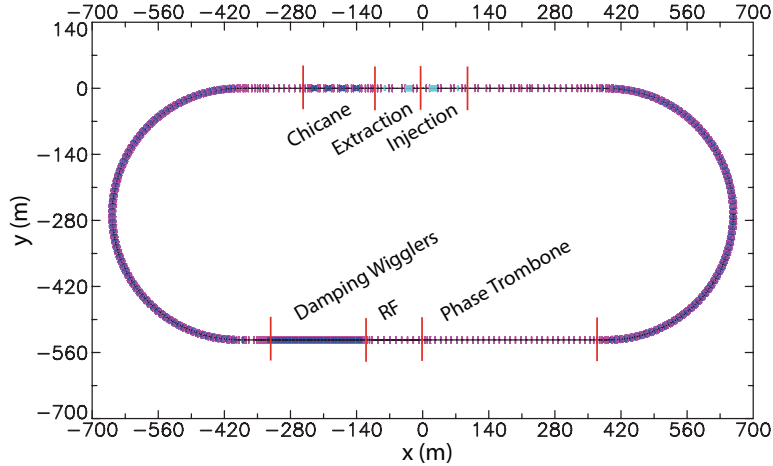}
        \caption{Layout of DTC04 lattice.}
        \label{fig:dtc04_layout}
    \end{figure}

\begin{figure*}[t]
    \centering
        \includegraphics[width=\textwidth]{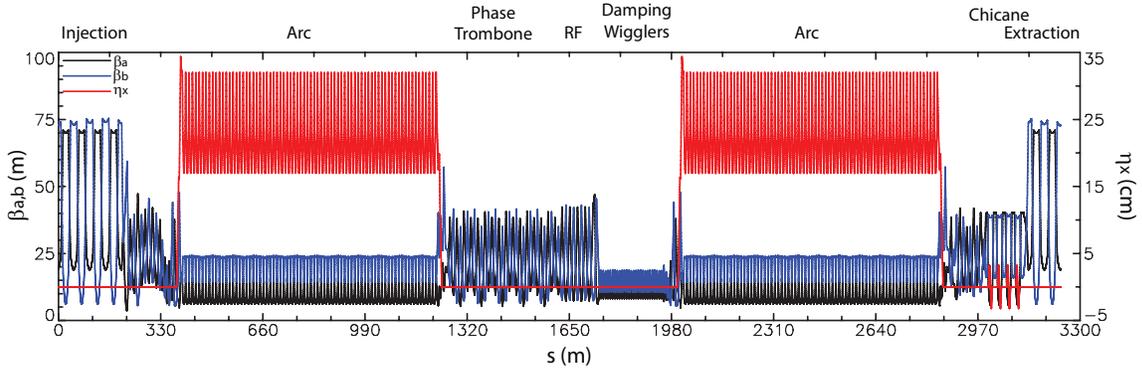}
        \caption{Horizontal and vertical $\beta$ functions and horizontal
        dispersion for the DTC04 lattice. The ripple in $\beta_b$
        in the arcs is due to aliasing, and is not physical.}
        \label{fig:dtc04_twiss}
    \end{figure*}

    The baseline ILC damping ring design is the DTC04 lattice
    developed by Rubin \emph{et al.} \cite{IWFLC12:Shanks}. The lattice is a
    3.2~km racetrack with a modified TME-type arc cell. The zero-dispersion
    straights are based on the work of Korostelev and Wolski
    \cite{IPAC10:WEPE096}. A schematic of the ring is shown in
    figure
    \ref{fig:dtc04_layout}. Optics functions are shown in figure
    \ref{fig:dtc04_twiss}. Tables \ref{tab:dtc04_summary} and
    \ref{tab:mag_count} summarize the lattice parameters and number of magnets
    of each class.
    The quoted emittances $\epsilon_{a,b}$ are the horizontal-like and
    vertical-like normal modes, respectively. In the absence of coupling,
    these will correspond to $\epsilon_{x,y}$. Decomposition into
    normal modes is discussed elsewhere \cite{PhysRevSTAB.9.024001,
    PhysRevLett.111.104801, PhysRevSTAB.15.124001}, and as such, a
    derivation is omitted here.

        \begin{table}[htb]
        \centering
            \caption {Summary of the DTC04 lattice parameters.}
            \label{tab:dtc04_summary}
            \begin{tabular}{ccc}
                \toprule[1pt]
                \addlinespace[2pt]
                \textbf{Parameter} & \textbf{Value} & \textbf{Units}\\
                \midrule[0.5pt]
                Circumference & 3239 & m\\
                Energy      & 5.0 & GeV \\
                Betatron Tunes ($Q_x$, $Q_y$)& (48.850, 26.787) & \\
                Tune Advance per Arc  & (16.418, 6.074) & \\
                Chromaticity ($\xi_x,\xi_y$) & (1.000, 0.302)   & \\
                Chromaticity per Arc  & (9.074, 10.896) & \\
                Train Repetition Rate & 5 & Hz \\
                Bunch Population      & $2\times 10^{10}$ &  \\
                Extracted $\epsilon_a^{geometric}$ & 0.6 & nm\\
                Extracted $\epsilon_b^{geometric}$  & $<2$ & pm\\
                Extracted Bunch Length  & 6  & mm \\
                Extracted $\sigma_E/E$ & 0.11 & \% \\
                Damping Time      & 24  & ms \\
                Wiggler $B^{max}$ & 1.5 & T \\
                \bottomrule[1pt]
            \end{tabular}
        \end{table}

    \begin{table}[htb]
        \centering
        \caption {Summary of elements in the DTC04 lattice.}
        \label{tab:mag_count}
        \begin{tabular}{cc}
            \toprule[1pt]
            \addlinespace[2pt]
            \textbf{Class} & \textbf{Count} \\
            \midrule[0.5pt]
            Beam Position Monitor  & 511 \\
            Dipole          & 164 \\ 
            Horizontal Steering & 150 \\
            Vertical Steering  & 150 \\
            Combined H+V Steering  & 263 \\
            Quadrupole      & 813 \\
            Skew Quadrupole & 160 \\
            Sextupole       & 600 \\
            Damping Wigglers & 54    \\
            \bottomrule[1pt]
        \end{tabular}
    \end{table}

\begin{figure}[tbh]
    \centering
        \includegraphics[width=4 in]{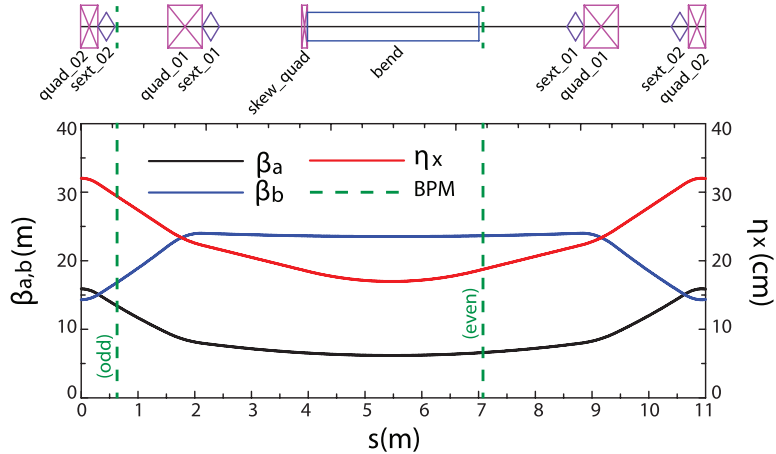}
        \caption{DTC04 arc cell. Two BPMs are available in
        each arc cell, and are either odd-indexed or even-indexed.}
        \label{fig:dtc04_arccell}
    \end{figure}

    The arc cell layout is shown in figure \ref{fig:dtc04_arccell}.
    Arc cells are comprised of one dipole, three quadrupoles, four
    sextupoles, one horizontal and one vertical steering corrector,
    one skew quadrupole, and two beam position monitors.

\section{Misalignment and Correction Procedure}
    The effects of misalignments, BPM errors, multipoles, and
    choice of correction procedure are evaluated by the program
    \texttt{ring\_ma2} \cite{PhysRevSTAB.17.044003}. The \texttt{ring\_ma2}
    workflow is structured as follows:

    \begin{enumerate}
        \item Assign random misalignments and BPM errors with user-defined
        amplitudes to the ideal lattice to create a realistic machine
        model.
        \item Simulate beam based measurements of optics functions
        including the effects of BPM measurement errors.
        \item Compute and apply corrections for each iteration
        based on the simulated measurements.
        \item After each correction iteration, record the
        effectiveness of the correction in terms of emittances
        and optics functions.
    \end{enumerate}

    The entire procedure is repeated typically 100 times, each with a
    randomly chosen distribution of errors in order to generate
    statistics sufficient for analysis. Misalignments are
    allowed on any element parameter described by \texttt{Bmad},
    including but not limited to: position offsets and angles,
    random field strength errors, and systematic and random multipole
    errors. BPM measurement errors allowed in the simulation
    include BPM-to-quadrupole transverse offset, tilt, button-by-button
    gain error, and button-by-button timing errors.

    The particular configuration of the guide field magnets, correctors,
    and BPMs is referred to as a scenario. For each scenario, 100 lattices
    are generated with random errors and misalignments. Each misaligned
    lattice is referred to as a seed. As noted above, typically 100 seeds
    are used for the evaluation of each scenario, and the same 100 seeds
    are used when evaluating multiple scenarios. That is, for a given seed,
    each element in a lattice for each scenario starts with the same
    misalignments and field errors, regardless of whether a subset of
    elements or detectors are vetoed or disabled during the correction.

    The method used for simulating measurements with BPM errors is
    based on the method developed for diagnosing emittance dilution at
    CesrTA \cite{PhysRevSTAB.17.044003}. In the original analysis, modeling
    the full correction procedure includes turn-by-turn tracking
    for several damping times in order to accurately simulate measurements
    of betatron phase and coupling. The turn-by-turn measurements are then
    post-processed identically to actual machine data into orbit, dispersion,
    or phase and coupling data. Though this method for simulation is as close
    to modeling the actual measurement process, it is very
    time-consuming. In the case of the ILC damping ring lattice,
    this method is prohibitively slow due
    to the large number of elements. Two modifications have been made to the
    measurement simulation procedure in order to improve the efficiency of
    the calculations such that the characterization method developed in
    \cite{PhysRevSTAB.17.044003} may be applied.

    First, there is a modification to tracking in the damping wigglers.
    When accounting for wiggler nonlinearities is required, a
    fully-symplectic Lie map is used for tracking through the
    damping wigglers \cite{PAC05:Crittenden, IPAC12:TUPPR065}.
    However, as it is presently implemented in \texttt{Bmad},
    this tracking method is computationally intensive. When the full
    nonlinearities of the damping wigglers are known to not affect the
    resulting simulation, a simplified MAD-style ``bend-drift-bend'' wiggler
    model which preserves the radiation integrals can be used to reduce the
    time required for simulations. Radiation integrals and
    horizontal emittances the two wiggler models are shown in
    table \ref{tab:wig_model_comparison}. The horizontal emittances computed
    with the two models differ by less than 3\%.  The level of agreement is
    more than sufficient for the studies presented here.

    \begin{table}[htb]
    \centering
        \caption {Comparison of emittance and radiation
        integrals for DTC04 lattice with two wiggler models: a
        fully-symplectic Lie map, and a simplified
        ``bend-drift-bend'' model.}
        \label{tab:wig_model_comparison}
        \begin{tabular}{cccc}
            \toprule[1pt]
            \addlinespace[2pt]
            \textbf{Parameter} & \textbf{Lie Map} & \textbf{Bend-Drift-Bend}
            & \textbf{$\Delta$} \\
            \midrule[0.5pt]
            $\epsilon_a$ & 0.553~nm & 0.567~nm & -2.51\%\\
            $I_1$ & 1.08 & 1.08 & 0.00\%\\
            $I_2$ & 0.512~m$^{-2}$ & 0.515~m$^{-2}$ & -0.59\%\\
            $I_3$ & $3.39\times 10^{-2}$~m$^{-3}$ &
            $3.40\times 10^{-2}$~m$^{-3}$ & -0.44\%\\
            $I_4$ & $1.98\times 10^{-4}$~m$^{-2}$ &
            $2.01\times 10^{-4}$~m$^{-2}$ &
            -1.82\%\\
            $I_5$ & $7.71 \times 10^{-6}$~m$^{-2}$ & $7.95 \times
            10^{-6}$~m$^{-2}$ & -3.13\%\\
            \bottomrule[1pt]
        \end{tabular}
    \end{table}

    Comparisons of simulations using the full wiggler model
    and the reduced MAD-style ``bend-drift-bend'' wiggler have shown that
    wiggler nonlinearities do not significantly affect the correction of
    optics functions or the final emittance. Therefore, the simplified
    ``bend-drift-bend'' wiggler model is used for all \texttt{ring\_ma2}
    studies on the ILC damping ring lattice. This reduces the time
    required to simulate the correction procedure for one
    randomly-misaligned lattice by a factor of 15.

    The second modification affects the process through which optics
    measurements are simulated. In \cite{PhysRevSTAB.17.044003},
    measurements are simulated on a turn-by-turn basis,
    applying BPM measurement errors on every turn. Simulating one betatron
    phase measurement requires particle tracking for over 40,000
    turns, applying simulated BPM measurement errors on every turn. Even
    with the simplified wiggler model, this is prohibitively slow on a
    lattice as large as DTC04. An alternative is used in these studies
    where BPM errors are applied directly to the \texttt{Bmad}-computed
    optics functions. Although not as rigorous as the full measurement
    simulation method, side-by-side comparisons of the two methods have
    shown minimal difference in the resulting optics functions. As a
    result, the amount of time to simulate the correction procedure for
    one randomly-misaligned and corrected lattice is further reduced
    by another factor of 25, allowing the characterization
    of each configuration to be completed in a reasonable amount of time.

    The optics correction procedure used in this
    characterization was developed at \mbox{CesrTA}
    \cite{PhysRevSTAB.17.044003}, and relies on the measurement and
    correction of the betatron phase and coupling
    \cite{PRSTAB3:092801} using beam position monitors (BPMs) with
    turn-by-turn capabilities \cite{IPAC10:MOPE089}. The BPM modules are
    capable of pre-processing phase data in parallel, therefore this method
    has the benefit that data acquisition time is roughly independent of
    the number of BPMs, and is entirely independent of the number of
    correctors.

    The correction procedure used here is as follows:

    \begin{enumerate}
        \item Measure the closed orbit. Fit the data using all
        826 steering correctors (150 each of dedicated horizontal
        and vertical steerings, and 263 combined horizontal/vertical
        correctors) and apply corrections to the lattice model.
        \item Measure the betatron phase/coupling and dispersion.
        Correct betatron phase and horizontal dispersion to the design values
        using all 813 normal quadrupoles; simultaneously correct betatron
        coupling using all 160 skew quadrupole correctors, and apply
        all corrections.
        \item Remeasure the closed orbit, coupling, and vertical dispersion;
        simultaneously correct all machine data using all 513 steering
        correctors and 160 skew quadrupole correctors and apply corrections.
    \end{enumerate}

    The correction procedure described above is routinely used at CesrTA,
    therefore a direct comparison of simulation with measurement is available
    to validate the emittance tuning method. Details of the CesrTA
    characterization are available in \cite{CornellU2013:PHD:JPShanks,
    PhysRevSTAB.17.044003}. Although CesrTA has
    not achieved a vertical emittance below 2~pm as required for the ILC
    damping rings, simulated optics functions after correction are in agreement
    with measurements of the optics at CesrTA, and demonstrate accuracy in the
    modeling of the correction procedure. In practice, corrections based on
    betatron phase and coupling have yielded coupling and dispersion on par
    with that achieved using orbit response matrix analysis at CesrTA
    \cite{PAC09:WE6PFP104}.

\section{Error Tolerance of DTC04}
    The magnet misalignments,
    guide field errors, multipole errors, and BPM measurement error
    tolerances specified in the ILC Technical Design Report are
    applied in order to demonstrate that the baseline design satisfies
    the requirements for achieving vertical emittance. As will be seen,
    the full complement of BPMs is more than sufficient to
    achieve the requisite vertical emittance.

    \subsection{Nominal Lattice and Errors}
    \label{subsec:ringma2_nominal}

        Misalignments and BPM measurement error tolerances are summarized in
        table \ref{tab:dtc04_misalignments} in the Appendix, and are
        based on errors used in previous studies \cite{IWFLC12:Shanks}. In
        addition, quadrupole $k_1$ and sextupole $k_2$ errors are now included.
        Multipole error coefficients are taken from measurements of PEP-II
        guide field multipole errors by Cai \cite{cai_multipoles},
        and are summarized in table
        \ref{tab:dtc04_multipoles} in the Appendix. These error
        amplitudes and multipole coefficients, along with the full
        complement of 511 BPMs, define the nominal ILC-DR scenario.

\begin{figure}[tbh]
        \centering
            \includegraphics[width=4 in]{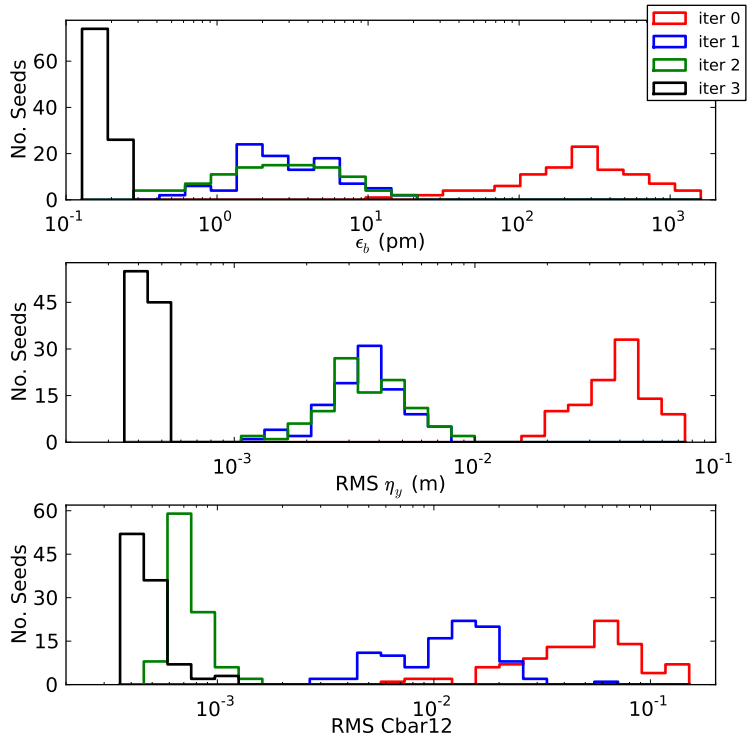}
            \caption{Distributions for emittance, dispersion, and coupling with
            misalignments, guide field multipole errors, and BPM errors as
            specified in the ILC TDR, using the full compliment of BPMs. Before
            correction (red), and after the first, second, and third
            corrections (blue, green, and black, respectively) are shown.
            Note that the horizontal axis is on a log-scale.}
            \label{fig:dtc04_nominal_ringma2}
        \end{figure}

        Results of \texttt{ring\_ma2} studies for this scenario are shown in
        figure \ref{fig:dtc04_nominal_ringma2} and are summarized in table
        \ref{tab:ringma2_corrections}. The coupling in model
        lattices is characterized using the $\bar{C}$ coupling matrix
        \cite{PRSTAB2:074001}, an extension of the Edwards and Teng formalism
        \cite{IEEETransNuclSci:EdwardsTeng}. In particular, the out-of-phase
        coupling matrix element $\bar{C}_{12}$ is used, as the $\bar{C}_{12}$
        measurement is insensitive to BPM rotation.

        \begin{table}[htb]
        \centering
            \caption {$95^{th}$-percentile vertical emittance,
            RMS vertical dispersion, and RMS coupling
            after each round of corrections, for the nominal lattice
            defined in Sec. }
            \label{tab:ringma2_corrections}
            \begin{tabular}{ccc}
                \toprule[1pt]
                \addlinespace[2pt]
                \textbf{Iteration}        & \textbf{Parameter} & $\bm{95^{th}}$
                \textbf{Percentile}\\
                \midrule[0.5pt]
                 Initial                  & $\epsilon_b$    & 1010~pm \\
                                          & RMS $\eta_y$        & 64.5~mm \\
                                          & RMS $\bar{C}_{12}$
                                          & $121\times 10^{-3}$ \\
                \midrule[0.1pt]
                 $x+y$                    & $\epsilon_b$    & 12.3~pm \\
                                          & RMS $\eta_y$        & 6.37~mm \\
                                          & RMS $\bar{C}_{12}$
                                          & $23.0\times 10^{-3}$\\
                \midrule[0.1pt]
                 $\phi_{a,b} + \bar{C}_{12} + \eta_x$       & $\epsilon_b$ & 10.9~pm \\
                                          & RMS $\eta_y$        & 6.92~mm \\
                                          & RMS $\bar{C}_{12}$
                                          & $1.11\times 10^{-3}$\\
                \midrule[0.1pt]
                 $y+\bar{C}_{12} + \eta_y$ & $\epsilon_b$   & 0.224~pm \\
                                           & RMS $\eta_y$       & 0.502~mm\\
                                           & RMS $\bar{C}_{12}$ &
                                           $0.704\times 10^{-3}$\\
                \bottomrule[1pt]
            \end{tabular}
        \end{table}

        Including misalignments, guide field errors, multipole errors, and BPM
        measurement error tolerances, 95\% of the resulting lattices have a
        vertical emittance below $\epsilon_b = 0.224~\textrm{pm}$
        after corrections. This is well below the emittance budget
        of $\epsilon_b = 2~\textrm{pm}$, though it is important to
        note that collective effects and
        other non-static sources of
        emittance dilution have not been accounted for in the
        calculation of the simulated emittance.

        The fundamental lower limit for the
        vertical emittance is determined by the finite opening angle of
        synchrotron radiation, and for the ILC damping rings is around 0.1~pm.
        The zero-current emittances achieved after correction for
        the nominal scenario is close to this limit,
        with 95\% of seeds well within a factor of three of the lower bound.

    \subsection{Reduced BPM Schemes}\label{subsec:reduced_bpms}

        It is clear that the baseline specifications for misalignments,
        multipole errors, and BPM and corrector distributions, as specified in
        the Technical Design Report, are more than sufficient to contain
        the static optics contribution to the vertical emittance
        within the 2~pm emittance budget. It is of interest to consider
        whether the number of BPMs can be reduced without compromising
        the effectiveness of the emittance correction procedure.

        The ILC damping ring lattice is wiggler-dominated,
        with $>80\%$ of synchrotron radiation being generated by the
        damping wigglers. As such, residual vertical dispersion in
        the wiggler straight will contribute significantly more to the
        vertical emittance than a comparable residual in the arcs.
        Preserving the 52 BPMs in the damping wiggler
        straight is therefore essential. The remaining 459 BPMs are
        either in other straight sections (159) or in the arcs
        (300).

        The baseline arc cell design has two BPMs, indicated in figure
        \ref{fig:dtc04_arccell}. The optics functions at the two
        BPMs are summarized in table \ref{tab:bpm_optics}.
        The betatron phase advance per arc is roughly 15 ($\approx 94.25$~rad),
        corresponding to about 10 BPMs per betatron wavelength in the
        arcs. The large number of BPMs and low phase advance per arc cell
        imply that not every BPM in the arcs may be necessary in order to
        maintain the ability to correct the lattice to below 2~pm vertical
        emittance.

        \begin{table}[htb]
        \centering
            \caption {Optics functions evaluated at each
            of the two periodic BPM locations in the DTC04 arc cell,
            denoted by ``odd-indexed'' and ``even-indexed''.}
            \label{tab:bpm_optics}
            \begin{tabular}{ccc}
                \toprule[1pt]
                \addlinespace[2pt]
                \textbf{Parameter} & \textbf{Odd-Indexed} &
                \textbf{Even-Indexed} \\
                \midrule[0.5pt]
                $\beta_a$          & 15.1~m & 6.5~m \\
                $\beta_b$          & 14.9~m & 23.5~m \\
                $\eta_a$           & 0.312~m & 0.185~m \\
                \bottomrule[1pt]
            \end{tabular}
        \end{table}

        Five scenarios for BPM distributions are examined. They
        are:

        \begin{enumerate}
            \item Full complement of BPMs, for a total of 511;
            10 BPMs per betatron wavelength. This is the nominal scenario
            from Sec. \ref{subsec:ringma2_nominal}.
            \item Remove odd-indexed arc BPMs, for a total of 361; 5
            BPMs per betatron wavelength, at locations with $\beta_b
            > \beta_a$
            \item Remove even-indexed arc BPMs, for a total of 361; 5
            BPMs per betatron wavelength, at locations with $\beta_b
            \approx \beta_a$
            \item Remove all odd-indexed and every other even-indexed arc BPM,
            for a total of 287; 2.5 BPMs per betatron wavelength,
            at locations with $\beta_b > \beta_a$
            \item Remove all even-indexed and every other odd-indexed arc BPM,
            for a total of 287; 2.5 BPMs per betatron wavelength, at locations
            with $\beta_b \approx \beta_a$
        \end{enumerate}

\begin{figure}[tbh]
        \centering
            \includegraphics[width=4 in]{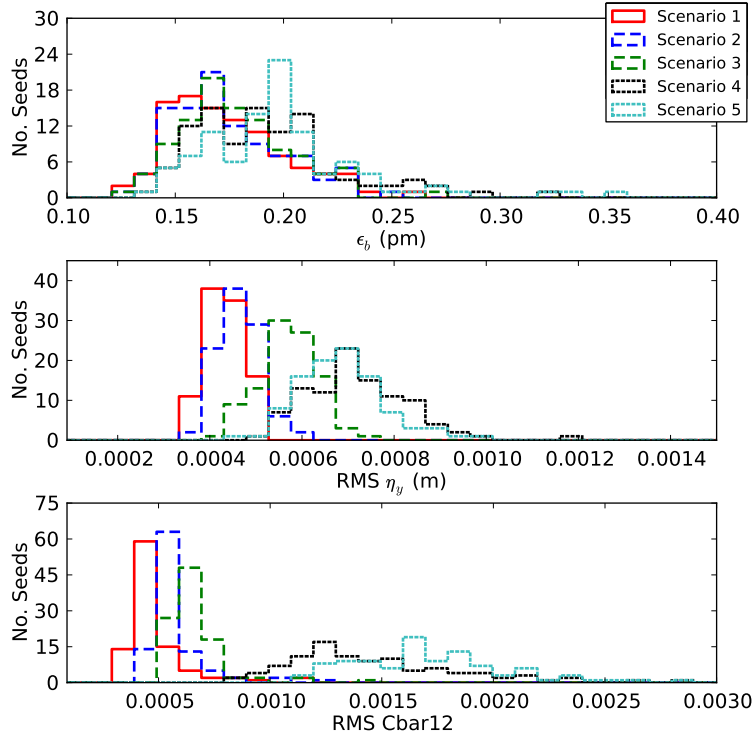}
            \caption{$\epsilon_b$, $\eta_y$, and $\bar{C}_{12}$ after
            the full correction procedure for various BPM distributions.
            See text for definitions of each scenario. Dashed lines indicate
            a scenario with 50\% of arc BPMs removed; dash-dotted lines
            indicate a scenario with 75\% of arc BPMs removed. Note that the
            horizontal axis is now linear.}
            \label{fig:dtc04_all_scenarios_ringma2}
        \end{figure}

        For each BPM distribution, the misalignment and correction
        procedure described in Sec. \ref{subsec:ringma2_nominal} is
        repeated. Magnet and BPM errors are identical for the five
        scenarios, therefore a direct comparison of resulting
        corrections is possible.
        Correction levels achieved for lattices with these BPM
        distributions are shown in figure \ref{fig:dtc04_all_scenarios_ringma2},
        and are summarized in table \ref{tab:ringma2_fewer_bpms}.

        \begin{table}[htb]
        \centering
            \caption {$95^{th}$ percentile of
            $\epsilon_b$, RMS $\eta_y$, and RMS $\bar{C}_{12}$
            after final correction
            with reduced BPM counts. See text for definitions of
            each BPM count scenario.}
            \label{tab:ringma2_fewer_bpms}
            \begin{tabular}{cccc}
                \toprule[1pt]
                \addlinespace[2pt]
                \textbf{BPM Scheme} & $\bm{\epsilon_b} \textbf{(pm)}$
                & \textbf{RMS} $\bm{\eta_y}$ \textbf{(mm)} & \textbf{RMS}
                $\bm{\bar{C}_{12}}$ \textbf{(\%)}\\
                \midrule[0.5pt]
                Scenario 1          & 0.224 & 0.502 & $0.0704$ \\
                Scenario 2          & 0.224 & 0.547 & $0.0790$ \\
                Scenario 3          & 0.225 & 0.670 & $0.0863$ \\
                Scenario 4          & 0.263 & 0.892 &  $0.216$ \\
                Scenario 5          & 0.269 & 0.838 &  $0.239$ \\
                \bottomrule[1pt]
            \end{tabular}
        \end{table}

        With half of the arc BPMs removed (scenarios 2-3),
        the correction procedure achieves vertical emittance below
        0.224-0.225pm for 95\% of the lattices.
        With the removal of 3/4 of BPMs in the arcs (scenarios 4-5),
        there is a slight increase in the $95^{th}$ percentile $\epsilon_b$
        to 0.263-0.269~pm, still well below the 2~pm requirement.
        There is a weak preference for retaining
        the ``even-indexed'' BPMs (scenarios 2 and 4) as compared to
        the ``odd-indexed'' BPMs (scenarios 3 and 5). Scenarios 3 and
        5 will therefore be omitted from further discussion.

    \subsection{Effect of Random BPM Failures}

        From the proceeding discussion, it is clear that the
        BPM count can be reduced substantially with respect to the
        baseline design specification without compromising the
        effectiveness of the emittance tuning procedure.
        However, this assumes all remaining BPMs are fully
        functional. This is rarely the case in a real machine, and
        as such, the effects of random BPM failures must be
        explored before deeming a BPM distribution acceptable.

        Again using the same random seeds as in previous
        simulations for generating lattice and BPM errors,
        a random subset of BPMs is flagged as not functional,
        and therefore not used in corrections. As there are multiple
        BPM distributions under evaluation, only the fraction of
        BPMs tagged as ``failed'' is constant between BPM distribution
        scenarios, rather than failing a fixed number of
        BPMs. The exact subset of BPMs for each test will therefore
        not remain the same between each BPM distribution scenario.
        It is not possible to have a truly random distribution of
        BPM failures while requiring that the same BPMs are tagged
        as ``failed'' between the three BPM distribution scenarios
        evaluated here, since the total number and distribution of BPMs varies
        from one scenario to the next. As such, BPM failure tests on each
        individual BPM distribution scenario are directly comparable
        seed-by-seed, but there is a small statistical variation
        between the different BPM distribution scenarios for a fixed
        percentage of BPM failures.

\begin{figure}[tbh]
        \centering
            \includegraphics[width=3.5 in]{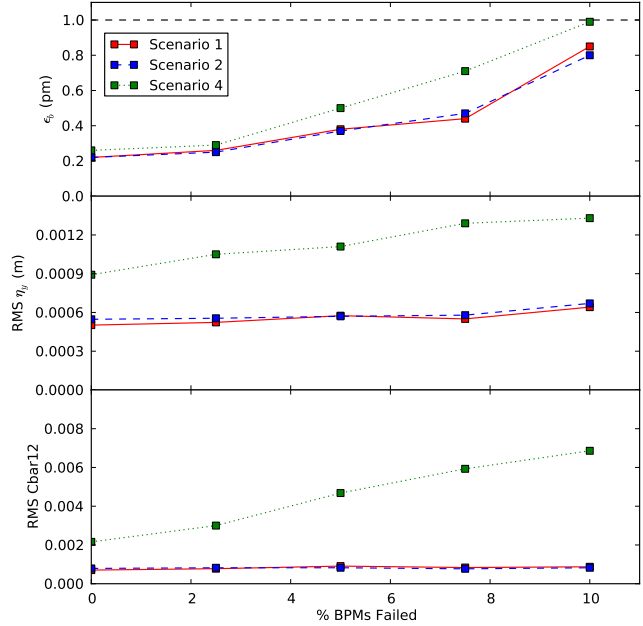}
            \caption{$95^{th}$-percentile $\epsilon_b$, RMS $\eta_y$,
            and RMS $\bar{C}_{12}$ as a function of the fraction
            of BPMs randomly disabled, for three potential BPM
            distributions. See text for definitions of each BPM
            distribution. The dashed black line indicates
            $\epsilon_b = 1$~pm.}
            \label{fig:ringma2_emit_vs_pct_failed}
        \end{figure}

        It is assumed that in practice, the damping rings would not
        be operated if more than 10\% of BPMs have failed.
        In this study, the fraction of BPMs tagged as
        ``failed'' is increased from 0\% to 10\%. Figure
        \ref{fig:ringma2_emit_vs_pct_failed} demonstrates the evolution of the
        $95^{th}$-percentile $\epsilon_b$, RMS $\eta_y$, and RMS $\bar{C}_{12}$
         with respect to the fraction of failed
        BPMs for the nominal case (scenario 1) and BPM distributions with $1/2$
        and $3/4$ of arc BPMs removed (scenarios 2 and 4, respectively).

        The resulting emittance, dispersion, and coupling of scenarios 1 and 2
        are nearly identical, within the expected statistical variation.
        Scenario 4, with only $1/4$ of arc BPMs with respect to the baseline
        design, shows more rapid growth in emittance, dispersion, and coupling
        as the fraction of failed BPMs increases. Vertical emittance growth
        from intra-beam scattering will increase with vertical dispersion
        \cite{PhysRevSTAB.16.104401}, thus making scenario 4
        unattractive. It is therefore concluded that one of the two
        BPMs in every arc cell may be removed without affecting the
        robustness of the emittance tuning procedure.

\section{Dynamic Aperture}
    After a lattice model has been misaligned and corrected using the
    simulated emittance tuning procedure, the dynamic aperture is evaluated
    through tracking. The dynamic aperture is defined as the maximum stable
    amplitude in the transverse plane. Trajectories with initial
    coordinates inside the dynamic aperture boundary will remain within
    that boundary for at least 1000 turns. The amplitude for trajectories
    with initial coordinates outside the boundary will be lost
    within 1000 turns. The tracking is repeated
    for off-energy particles to compute the energy dependence of the
    dynamic aperture.

    The full wiggler map is required for evaluating the dynamic aperture in
    order to account for wiggler nonlinearities.
    This greatly increases the required computation time for a
    dynamic aperture study, however individual jobs are easily parallelized.
    Analysis has shown minimal variation of dynamic
    aperture between different sets of misalignments and corrections, given
    the same misalignment amplitudes. Therefore only one seed for each test
    configuration is evaluated for dynamic aperture. Here, the
    random seed corresponding to the $95^{th}$-percentile lattice after
    misalignments and correction is used.

\begin{figure}[ht]
    \centering
    \begin{subfigure}[t]{0.45\textwidth}
        \centering
        \includegraphics[width=2.75 in]{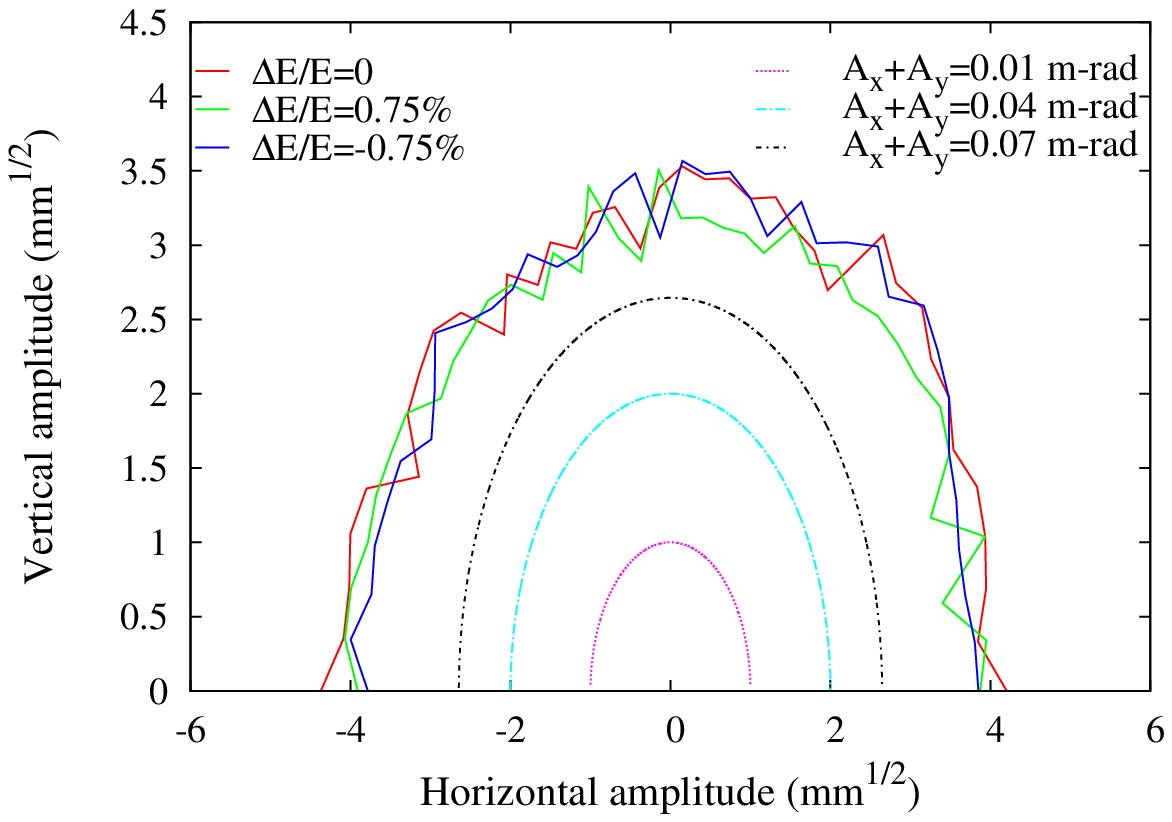}
        \caption{Nominal scenario as
        defined in Sec. \ref{subsec:ringma2_nominal}.}
        \label{fig:dtc04_da_nominal_1xmult}
    \end{subfigure}
    \quad
    \begin{subfigure}[t]{0.45\textwidth}
        \centering
        \includegraphics[width=2.75 in]{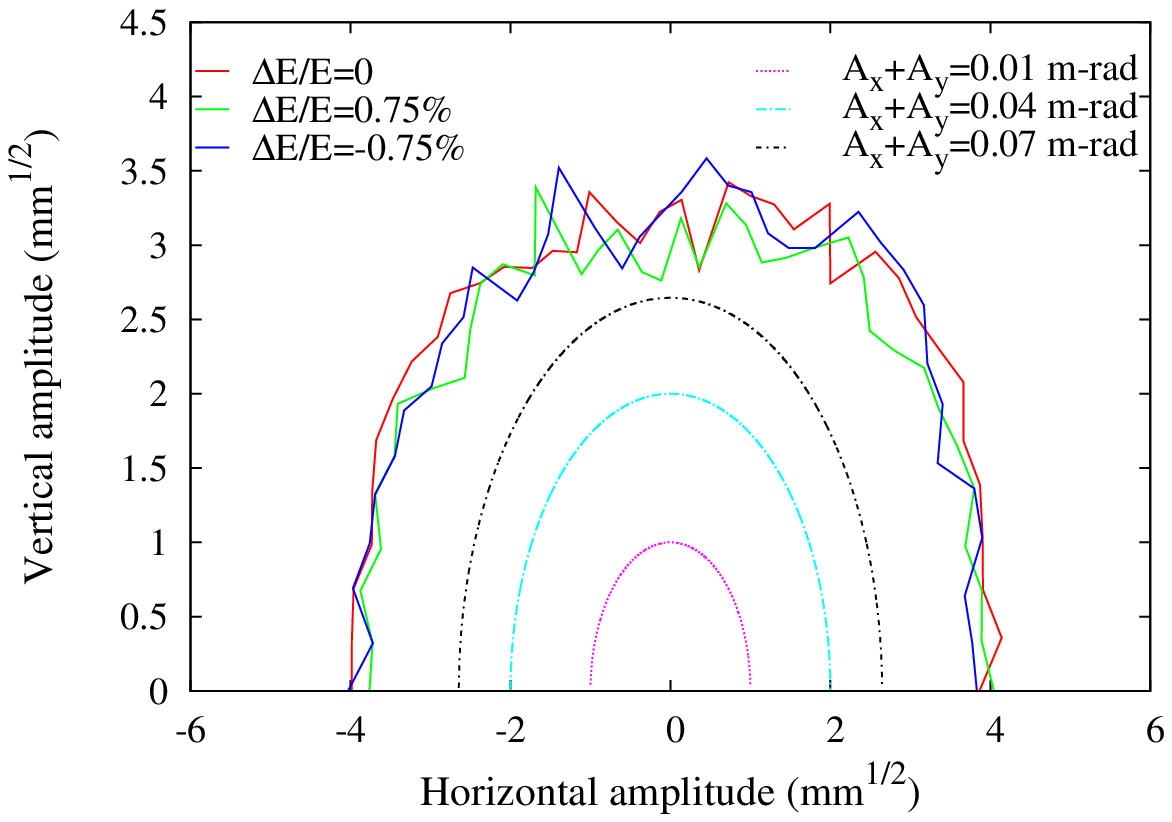}
        \caption{Multipole
        errors increased by a factor of 5x.}
        \label{fig:dtc04_da_nominal_5xmult}
    \end{subfigure}
    \quad
    \begin{subfigure}[t]{0.45\textwidth}
        \centering
        \includegraphics[width=2.75 in]{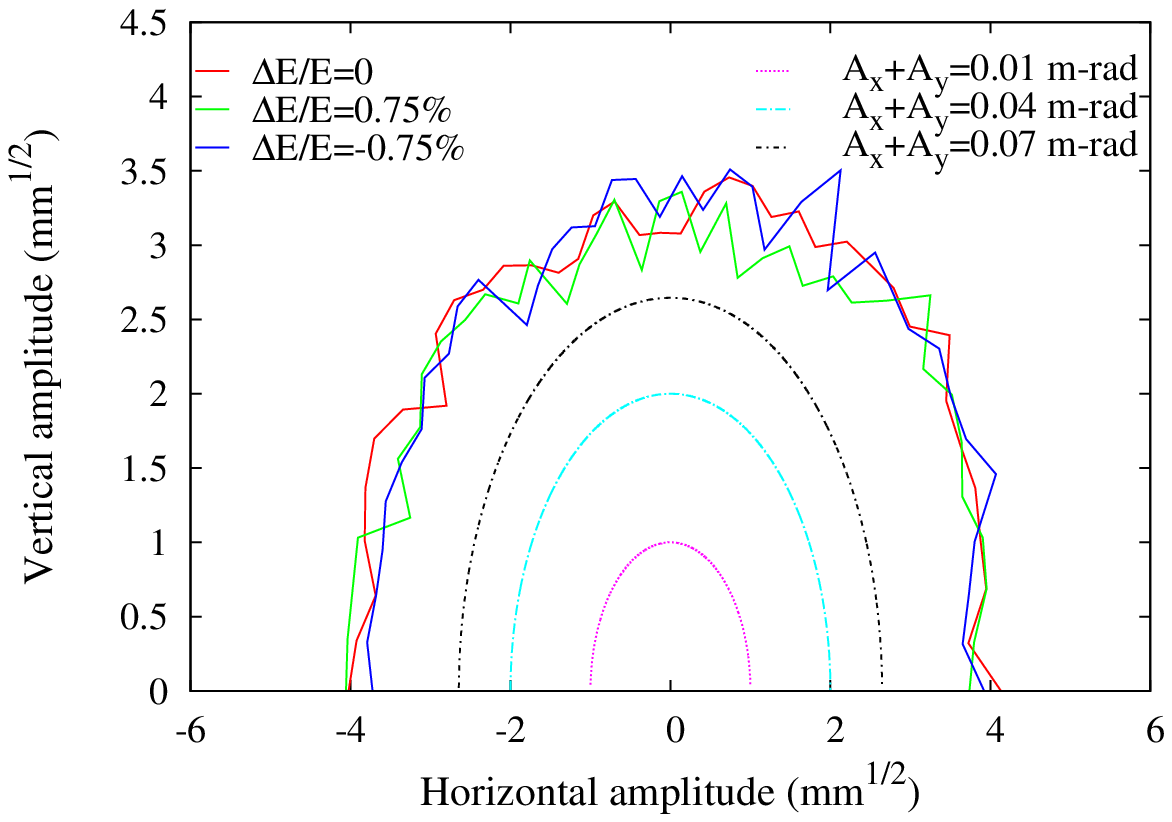}
        \caption{Multipole
        errors increased by a factor of 10x.}
        \label{fig:dtc04_da_nominal_10xmult}
    \end{subfigure}
    \quad
    \begin{subfigure}[t]{0.45\textwidth}
        \centering
        \includegraphics[width=2.75 in]{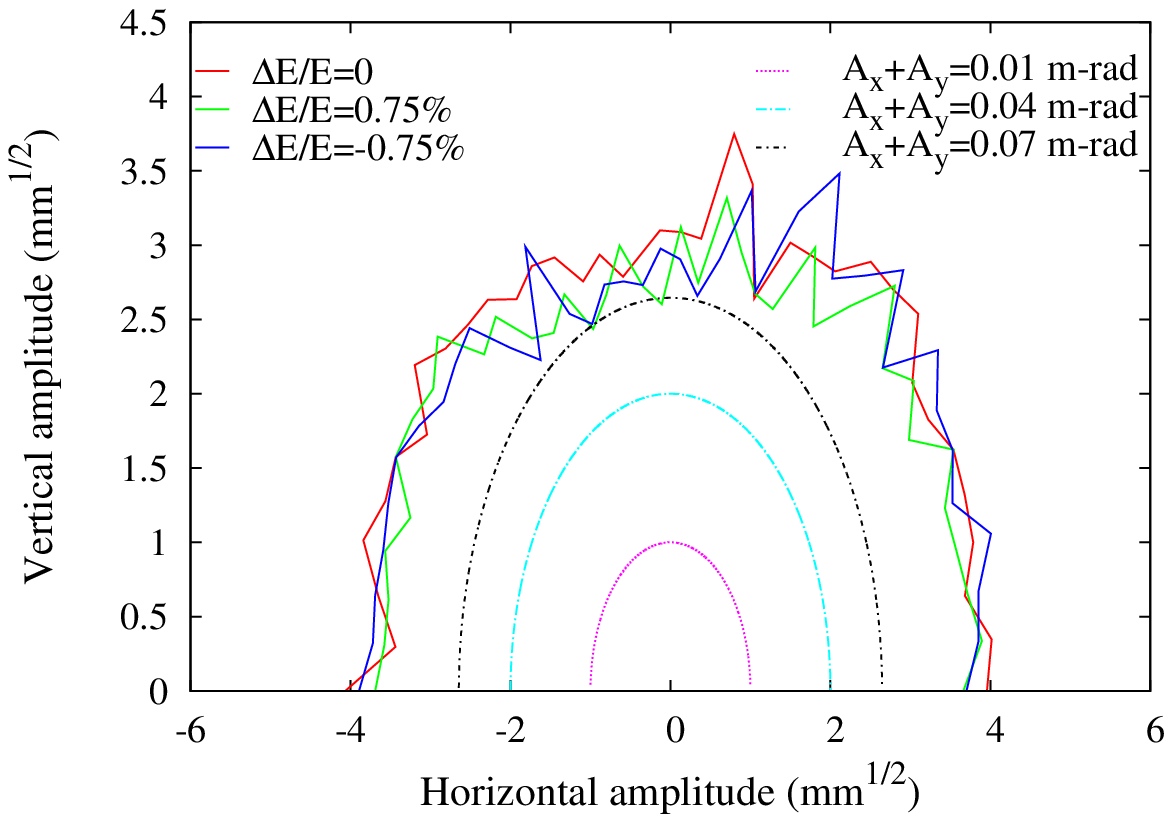}
        \caption{Multipole
        errors increased by a factor of 20x.}
        \label{fig:dtc04_da_nominal_20xmult}
    \end{subfigure}
    \caption{Dynamic aperture for the DTC04 lattice, while varying
    the amplitude of systematic and random multipole errors. All
    scenarios shown include the full complement of BPMs.
    The overlayed black ellipse represents the maximum expected
    amplitude for bunches
        transferred from the positron source \cite{ilc:tdr}.}
    \end{figure}

    Figure \ref{fig:dtc04_da_nominal_1xmult} shows the
    dynamic aperture for the nominal scenario, as defined in
    Sec. \ref{subsec:ringma2_nominal}.
    It is evident that the systematic and random multipole
    errors as specified in the ILC Technical Design Report are more
    than sufficient to achieve the required dynamic aperture for
    accepting an incoming bunch from the positron source.

    The possibility that magnet manufacturing requirements
    may be relaxed by increasing the tolerance on
    magnet multipole errors is explored.
    Starting with the nominal DTC04 lattice
    characterized in Section \ref{subsec:ringma2_nominal}, both systematic and random
    multipole coefficients are increased by a constant multiplier from 5x-20x
    with respect
    to the values in table \ref{tab:dtc04_multipoles}. All other
    misalignments and BPM measurement errors are identical to the nominal
    scenario. Results are shown in figures
    \ref{fig:dtc04_da_nominal_5xmult}--\ref{fig:dtc04_da_nominal_20xmult}.

%
%

    The multipole error constraints as specified in the ILC TDR are
    evidently over-specified by a factor of 20.

\section{Summary and Future Work}
    A method has been presented for determining the
    effectiveness of corrections for the ILC damping rings and the
    resulting dynamic aperture after correction, using
    the emittance tuning algorithm developed and used at CesrTA. Lattice models
    include misaligned magnets, magnet strength and multipole errors, and BPM
    measurement errors. The method is versatile and may be applied to any
    closed lattice to evaluate the sensitivity of the effectiveness of
    emittance corrections to changes in BPM and corrector distributions,
    multipole tolerances, and more.

    Based on this method, the ILC damping ring design, misalignment
    tolerances, and BPM measurement errors as presented in the ILC
    Technical Design Report (TDR) have been shown
    to achieve the required zero-current vertical emittance and dynamic
    aperture. Note that collective effects such as intra-beam scattering,
    electron cloud, and fast-ion instability can only increase the vertical
    emittance above this minimum.

    One of the two BPMs in every arc cell may be removed without impact on the
    correction capabilities, even when accounting for up to a 10\%
    BPM failure rate. This leaves approximately 5 BPMs per betatron wavelength
    in the arcs. A further reduction of BPMs in the arcs to 1/4
    of the original count ($\approx$2.5 BPMs per betatron wavelength)
    causes a substantial increase in
    dispersion and coupling, subsequently fueling emittance growth
    from collective effects.
    Specifications for magnet multipole errors can be relaxed
    by a factor of twenty without compromising the dynamic aperture.

    An infrastructure exists for evaluating further lattice modifications,
    such as sextupole optimizations, increased magnet or BPM errors,
    changes in corrector distributions, or a change in working point.
    Further lattice developments in these areas can therefore be easily
    evaluated for the ILC damping rings or any arbitrary closed
    lattice.

\begin{acknowledgments}
    This work was supported by the National Science Foundation grant
    PHY-1002467 and Department of Energy grant DE-SC0006505.
\end{acknowledgments}

\appendix

\setcounter{secnumdepth}{0}

\section{Appendix: DTC04 Magnet Misalignments and Multipoles}
\label{sec:appendix}

    Multipole coefficients are defined in the following way:

    \begin{equation}
        \left(B_y + \imath B_x\right) = B(r) \sum_{n=1}(b_n + \imath
        a_n) \left(\frac{x}{r} + \imath\frac{y}{r}\right)^{n-1}
    \end{equation}

    \noindent where n is the multipole order, and $b_n, a_n$
    are the normal and skew components, respectively. Multipoles are
    evaluated at a reference radius of $r=3$~cm for dipoles, 5~cm
    for quadrupoles, and 3.2~cm for sextupoles.

        \begin{table}[htb]
        \centering
        \caption [Errors used in ILC-DR studies]
            {Misalignments and errors introduced into the
            model ILC-DR lattice.}\label{tab:dtc04_misalignments}
            \begin{tabular}{rrrc}
                \toprule[1pt]
                \addlinespace[2pt]
                \textbf{Element} & \textbf{Error} & \textbf{Amplitude} &
                \textbf{Units} \\
                \midrule[0.5pt]
                Dipole     & Roll                & 50 & $\mu$rad   \\
                \midrule[0.5pt]
                Quadrupole & $x$, $y$ Offset          & 25 & $\mu$m \\
                           & Tilt                & 50 & $\mu$rad   \\
                           & k1                  & 0.1\% & \%\\
                \midrule[0.5pt]
                Sextupole  & $x$, $y$ Offset          & 25 & $\mu$m   \\
                           & Tilt                & 25 & $\mu$rad   \\
                           & k2                  & 1\%& \%   \\
                \midrule[0.5pt]
                Wiggler    & Tilt                & 100&$\mu$rad   \\
                           & $y$ Offset            & 100&$\mu$m   \\
                \midrule[0.5pt]
                BPM & Diff. Resolution           & 1  &$\mu$m  \\
                    & Abs. Resolution            & 50 &$\mu$m   \\
                    & Tilt                       & 10 &mrad   \\
                    & Button Gains               & 0.5\%&\% \\
                    & Button Timing              & 10 &ps   \\
                \bottomrule[1pt]
            \end{tabular}
        \end{table}

        \begin{table}[htb]
        \centering
            \caption [ILC-DR multipoles for simulation]
            {Nominal multipoles (systematic and random) introduced into
            the model lattice. Coefficients are taken from multipole
            measurements by Y. Cai at PEP-II
            \cite{cai_multipoles}.}\label{tab:dtc04_multipoles}
            \begin{tabular}{rcrr}
                \toprule[1pt]
                \addlinespace[2pt]
                \textbf{Element}  & \textbf{Multipole}    &
                \textbf{Systematic} & \textbf{Random} \\
                \midrule[0.5pt]
                Dipole      & b3 & $1.6\times10^{-4}$    &
                $8\times 10^{-5}$    \\
                            & b4 & $-1.6\times10^{-5}$   &
                            $8\times 10^{-6}$    \\
                            & b5 & $7.6\times10^{-5}$    &
                            $3.8\times 10^{-5}$  \\
                \midrule[0.5pt]
                Quadrupole  & a3 & $-1.15\times10^{-5 }$ &
                $7.25\times 10^{-5}$ \\
                            & a4 & $ 1.41\times10^{-5 }$ &
                            $1.27\times 10^{-4}$ \\
                            & a5 & $ 6.2\times10^{-7}  $ &
                            $1.62\times 10^{-5}$ \\
                            & a6 & $-4.93\times10^{-5 }$ &
                            $3.63\times 10^{-4}$ \\
                            & a7 & $-1.02\times10^{-6 }$ &
                            $6.6\times 10^{-6}$  \\
                            & a8 & $ 3.8\times10^{-7  }$ &
                            $6.6\times 10^{-6}$  \\
                            & a9 & $-2.8\times10^{-7  }$ &
                            $4.9\times 10^{-6}$  \\
                            & a10& $-5.77\times10^{-5 }$ &
                            $2.33\times 10^{-4}$ \\
                            & a11& $-3.8\times10^{-7  }$ &
                            $3.5\times 10^{-6}$  \\
                            & a12& $-6.53\times10^{-6 }$ &
                            $3.66\times 10^{-5}$ \\
                            & a13& $ 1.2\times10^{-6  }$ &
                            $8.6\times 10^{-6}$  \\
                            & a14& $-7.4\times10^{-7  }$ &
                            $4.46\times 10^{-5}$ \\
                            & b3 & $-1.24\times10^{-5 }$ &
                            $7.61\times 10^{-5}$ \\
                            & b4 & $ 2.3\times10^{-6  }$ &
                            $1.32\times 10^{-4}$ \\
                            & b5 & $-4.3\times10^{-6  }$ &
                            $1.5\times 10^{-5}$  \\
                            & b6 & $ 3.4\times10^{-4  }$ &
                            $1.65\times 10^{-4}$ \\
                            & b7 & $ 3\times10^{-7    }$ &
                            $6.7\times 10^{-6}$  \\
                            & b8 & $ 6\times10^{-7    }$ &
                            $8.9\times 10^{-6}$  \\
                            & b9 & $ 6\times10^{-7    }$ &
                            $4.6\times 10^{-6}$  \\
                            & b10& $-6.17\times10^{-5 }$ &
                            $2.46\times 10^{-4}$ \\
                            & b11& $-2\times10^{-7    }$ &
                            $4.2\times 10^{-6}$  \\
                            & b12& $ 3.6\times10^{-6  }$ &
                            $3.48\times 10^{-5}$ \\
                            & b13& $ 6\times10^{-7    }$ &
                            $9.2\times 10^{-6}$  \\
                            & b14& $ 1\times10^{-6    }$ &
                            $4.76\times 10^{-5}$ \\
                \midrule[0.5pt]
                Sextupole   & b4 & $1\times 10^{-4}$     & $1\times 10^{-4}$ \\
                            & b5 & $5\times 10^{-5}$     & $3\times 10^{-5}$ \\
                            & b6 & $3.5\times 10^{-4}$   & $1\times 10^{-4}$ \\
                            & b7 & $5\times 10^{-5}$     & $3\times 10^{-5}$ \\
                            & b8 & $5\times 10^{-5}$     & $3\times 10^{-5}$ \\
                            & b9 & $5\times 10^{-5}$     & $3\times 10^{-5}$ \\
                            & b10& $5\times 10^{-5}$     & $3\times 10^{-5}$ \\
                            & b11& $5\times 10^{-5}$     & $3\times 10^{-5}$ \\
                            & b12& $1.6\times 10^{-3}$   & $1\times 10^{-4}$ \\
                            & b13& $5\times 10^{-5}$     & $3\times 10^{-5}$ \\
                            & b14& $5\times 10^{-5}$     & $3\times 10^{-5}$ \\
                \bottomrule[1pt]
            \end{tabular}
        \end{table}

\clearpage

\bibliography{CesrTA}

\end{document}